\documentclass[prl,aps,showpacs,groupedaddress,superscriptaddress,twocolumn]{revtex4-1}
\usepackage{amsmath,amssymb}
\usepackage[utf8x]{inputenc}
\usepackage[english]{babel}
\usepackage{graphicx,epsfig}
\usepackage{todonotes}
\usepackage{hyperref}

\renewcommand{\vec}{\mathbf}

 \global\long\def\abs#1{\left|#1\right|}

  \global\long\def\v#1{\vec{#1}}

\usepackage{braket}
\usepackage{soul}
\usepackage{color}
\makeatother
\usepackage{placeins}

\begin{document}

\title{Dynamical quantum Cherenkov transition of fast impurities in quantum liquids}

\author{Kushal Seetharam}
\affiliation{Department of Electrical Engineering, Massachusetts Institute of Technology, Cambridge, Massachusetts 02139, USA}
\affiliation{Department of Physics, Harvard University, Cambridge, Massachusetts 02138, USA}

\author{Yulia Shchadilova}
\affiliation{Department of Physics, Harvard University, Cambridge, Massachusetts 02138, USA}

\author{Fabian Grusdt}
\affiliation{Department of Physics and Arnold Sommerfeld Center for Theoretical Physics (ASC), Ludwig-Maximilians-Universit\"at M\"unchen, Theresienstr. 37, M\"unchen D-80333, Germany}
\affiliation{Munich Center for Quantum Science and Technology (MCQST), Schellingstr. 4, D-80799 M\"unchen, Germany}

\author{Mikhail B. Zvonarev}
\affiliation{Universit\'e Paris-Saclay, CNRS, LPTMS, 91405, Orsay, France}
\affiliation{Russian Quantum Center, Skolkovo, Moscow 143025, Russia}
\affiliation{St. Petersburg Department of V.A. Steklov Mathematical Institute of Russian Academy of Sciences, Fontanka 27, St. Petersburg, 191023, Russia}

\author{Eugene Demler}
\affiliation{Institute for Theoretical Physics, ETH Z{\"u}rich, 8093 Z{\"u}rich, Switzerland}


\begin{abstract}

The challenge of understanding the dynamics of a mobile impurity in an interacting quantum many-body medium comes from the necessity of including entanglement between the impurity and excited states of the environment in a wide range of energy scales. In this paper, we investigate the motion of a finite mass impurity injected into a three-dimensional quantum Bose fluid as it starts shedding Bogoliubov excitations. We uncover a transition in the dynamics as the impurity's velocity crosses a critical value which depends on the strength of the interaction between the impurity and bosons as well as the impurity's recoil energy. We find that in injection experiments, the two regimes differ not only in the character of the impurity velocity abatement, but also exhibit qualitative differences in the Loschmidt echo, density ripples excited in the BEC, and momentum distribution of scattered bosonic particles. The transition is a manifestation of a dynamical quantum Cherenkov effect, and should be experimentally observable with ultracold atoms using Ramsey interferometry, RF spectroscopy, absorption imaging, and time-of-flight imaging.

\end{abstract}

\newpage

\maketitle

The nontrivial dependence of frictional forces on the velocity of moving objects is common to many classical systems, from sliding friction between solid bodies to drag forces in hydrodynamics. Electrodynamics provides an even more striking example, where a charged particle moving through a medium dissipates energy through light emission only if it’s velocity is above the speed of light in the medium, a phenomenon known as Cherenkov effect~\cite{Bolotovskii_2009}. A similar effect in quantum mechanical systems can be found when a particle travels through a superfluid. When the impurity particle is moving at a constant velocity, as is the case when it is so heavy that its recoil can be neglected, the Landau criterion states that the particle will only dissipate energy by generating excitations in the medium if it is traveling above the speed of sound of the superfluid. An impurity with finite mass, however, should exhibit dynamics that is far more complex.

Recently, ensembles of ultracold atoms have emerged as a versatile platform that is well suited to studying the physics of impurities, enabling dynamical control of their interaction strength and momenta. In a one-dimensional quantum gas, for example, polaronic renormalization of the impurity mass has been studied in Ref.~\cite{Catani2012}. Additionally, a mobile impurity pulled through a one-dimensional quantum gas by gravity was shown to exhibit Bloch oscillations on top of a finite drift velocity in the absence of a periodic lattice~\cite{meinert2017bloch}. In a three-dimensional Bose gas cooled into a Bose-Einstein Condensate (BEC), the radio-frequency (RF) spectrum~\cite{Hu2016,Jorgensen2016,Yan2019} and Loschmidt echo~\cite{Skou2021} of the polaron quasiparticle formed by the impurity and BEC has been probed as a function of interaction strength, but experiments have yet to examine quantities that are sensitive to the finite momentum of the impurity. For an infinite mass object, the speed of sound of the superfluid BEC sets a kinematic scale according to the Landau criterion, with a conical wavefront of Bogoliubov excitations emitted as the relative motion of the system exceeds this velocity~\cite{Astrakharchik_2004,Carusotto2006,El_2007,Gladush_2007,Gladush_2008}. A finite mass impurity, however, would recoil due to interactions with the surrounding quantum gas, yielding novel physics beyond the kinematic picture. Quantum fluctuations become highly relevant to the dynamics of even slowly moving impurities with finite mass~\cite{Grusdt2018}.


In this work, we investigate the interaction- and momentum-dependent behavior of a finite mass impurity moving through a three-dimensional BEC, focusing on the case of negative impurity-BEC scattering lengths. We examine (i) the lowest energy state of the system at finite momentum, which we refer to as the finite momentum ground state (FMGS), and (ii) the stationary state following temporal evolution of the impurity in an initial plane-wave state after interactions with the gas are quenched on. In both scenarios, we find that the impurity-gas interaction determines a critical system momentum, $\vec{P}_\textrm{crit}$, upon which the system exhibits a transition between two qualitatively different regimes. We observe that $\vec{P}_\textrm{crit}$ coincides for the FMGS and dynamical quench transitions. At small and intermediate interaction strengths, $\vec{P}_\textrm{crit}$ is set by the effective mass of the polaron. The recoil energy scale vanishes in the infinite impurity mass limit, where we recover the Landau criterion with $P_{\rm crit}=\abs{\vec{P}_\textrm{crit}}$ equivalent to the impurity mass times the BEC's speed of sound. 

\begin{figure*}[t!]
	\centering
	\includegraphics[width=0.92\textwidth]{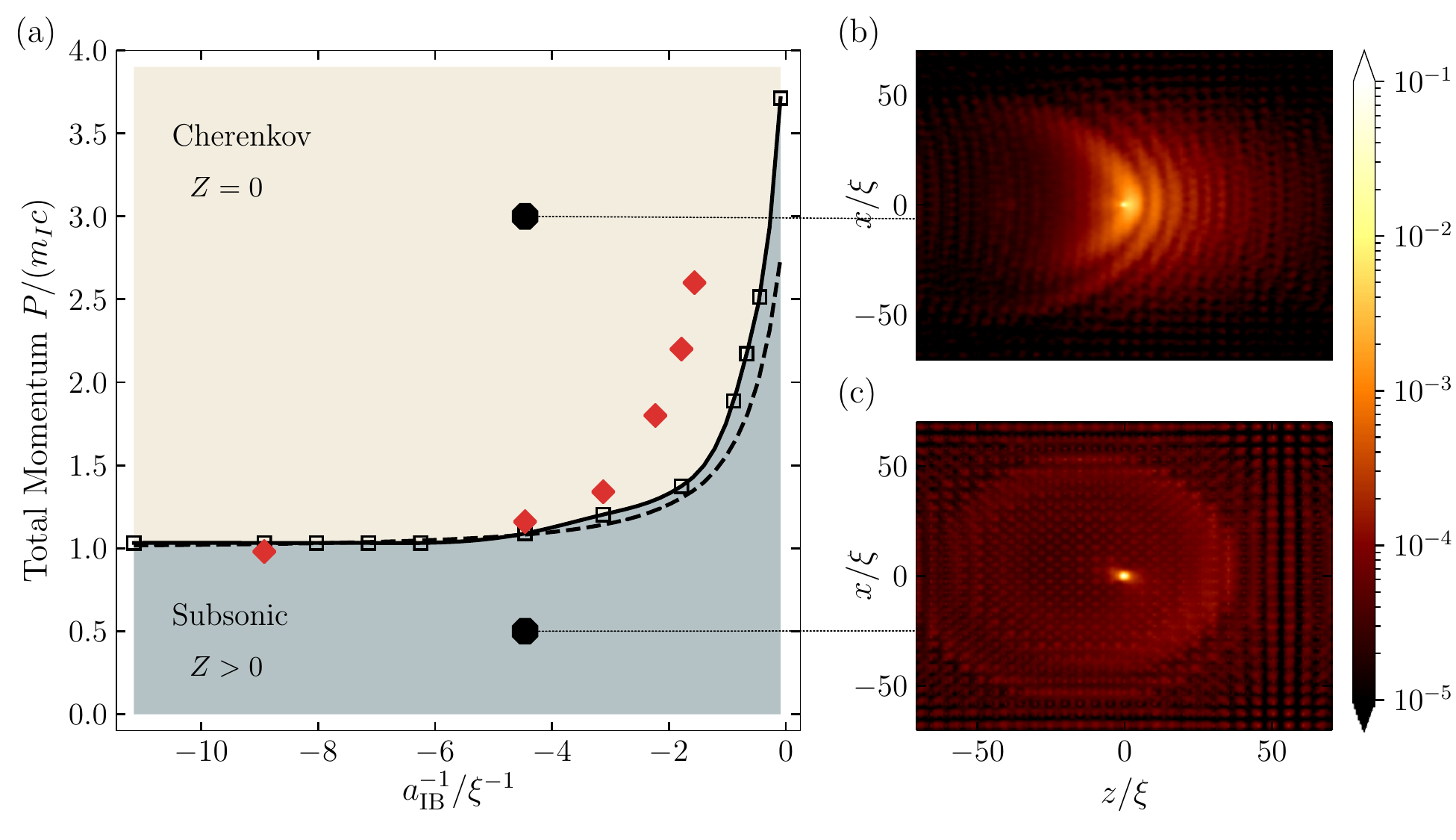}
	\caption{\textbf{Quantum Cherenkov transition of a mobile impurity interacting with a 3D BEC. (a)} Phase diagram depicting subsonic and Cherenkov regimes. The black squares mark the critical total system momentum, $P_\textrm{crit}$, numerically extracted from the discontinuity in the second derivative of the FMGS energy. The black solid line is an interpolated guideline for the transition. The black dashed line depicts $m^{*}c$, where $m^{*}$ is the polaron's mass, and $c$ the BEC's speed of sound. For weak and intermediate interactions, the dashed and solid lines coincide. The red diamonds show the numerically extracted transition points for the long time limit of the dynamical protocol. Panels \textbf{(b)} and \textbf{(c)} illustrate the real-space density distribution of atoms in the host liquid in each regime. The distributions are plotted at time $t=40\,\xi/c$ in the frame of the impurity propagating in the $z$-direction. The impurity-boson mass ratio is $m_{I}/m_{B}=1$.}
	\label{fig:GS_PhaseDiagram}
\end{figure*}

Below $\vec{P}_\textrm{crit}$, the polaron state overlaps with the free impurity and the impurity travels at an average velocity slower than the BEC's speed of sound, with this velocity depending on the momentum of the system. 
Above $\vec{P}_\textrm{crit}$, the polaron state is orthogonal to the free impurity and the impurity travels at the speed of sound with the rest of the system's momentum carried by long wavelength Bogoliubov excitations. In the injection experiment involving such a fast impurity, we find a shock wave and wake in the density of the host liquid, with this modulation traveling along with the impurity. In comparison to previously studied shock waves in superfluids generated by constant velocity heavy obstacles~\cite{Carusotto2006} or density defects~\cite{Dutton2001,Hoefer2006,Wan2006}, the dynamics of the density cone we observe is modified by the entanglement between the impurity and host atoms; this entanglement is included in the theoretical treatment of the system via the Lee-Low-Pines transformation~\cite{Lee1953} and results in additional interaction between Bogoliubov excitations (see~\cite{KS_CherenkovLetter_SM} for details). We therefore call this density modulation a polaron shock wave.

The finite momentum quantum transition we observe draws parallels to the classical Cherenkov effect, in that the impurity injected into a medium above a medium-dependent critical velocity saturates to a finite universal speed at late times while generating a cone of excitations in the medium.

\begin{figure*}[t!]
	\centering
	\includegraphics[width=0.91\textwidth]{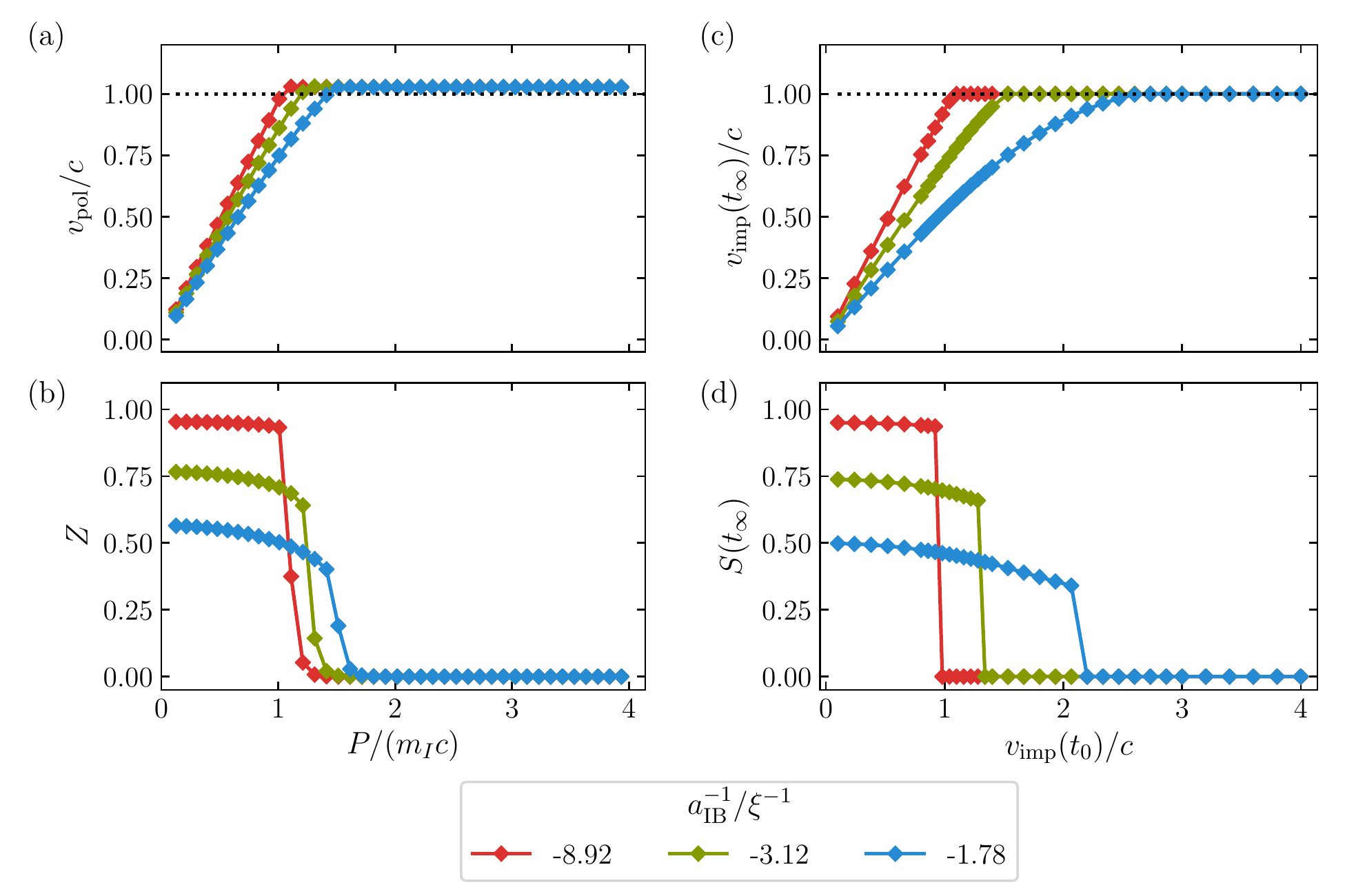}
	\caption{\textbf{FMGS and dynamical observables that witness the quantum Cherenkov transition.} The initial impurity velocity, $v_{\mathrm{imp}}(t_{0})$, corresponds to the total momentum of the system, $P$, which is conserved. \textbf{(a)} FMGS polaron group velocity. \textbf{(b)} FMGS quasiparticle residue. \textbf{(c)} Long time limit of the average impurity velocity. \textbf{(d)} Long time limit of the Loschmidt echo. The impurity-boson mass ratio is $m_{I}/m_{B}=1$.}
	\label{fig:ObsComp}
\end{figure*}


\textit{Results}.----We investigate an impurity immersed in a three-dimensional BEC where the total momentum of the system, $\vec{P}$, is conserved. The system is spherically symmetric, and $P=\abs{\v P}$ represents the magnitude of the total momentum. We treat the BEC using standard Bogoliubov theory. The phonon excitations of the bath have a dispersion that is quadratic at large momenta, $\omega_k \simeq k^2/(2m_{B})$, and linear at low momenta, $\omega_k \simeq ck$, where $k=\abs{\vec{k}}$. Here, $c=\sqrt{g_\textrm{BB}n_{0}/m_{B}}$ is the BEC's speed of sound, $m_{B}$ is the boson mass, $g_{BB}$ is the boson-boson interaction strength, and $n_{0}$ is the condensate density. The interaction between the impurity and BEC is parameterized by the impurity-boson scattering length $a_\textrm{IB}$, which we express in units of the condensate healing length $\xi = (2m_{B} g_\mathrm{BB} n_{0})^{-1/2}$. We use an interaction Hamiltonian that keeps terms that are relevant for the strong coupling regime. The Hamiltonian model and variational wavefunction used to derive equations of motion are the same as in Ref.~\cite{Shchadilova2016}, and in this work we tackle the numerical challenge of solving the equations of motion with enough resolution to accurately capture the physics of the system. The methods are further discussed in~\cite{KS_CherenkovLetter_SM}, with more detail provided in Ref.~\cite{Seetharam2021_CherenkovLong}. In this work, we examine the case of negative impurity-BEC scattering lengths. The dynamics of impurities with positive impurity-BEC scattering lengths can be affected by the formation of multi-particle bound states~\cite{Shchadilova2016,Drescher2019}, and requires a separate analysis.

We now elaborate how various observables both in the FMGS and at long times after a dynamical quench show signs of the transition as the total system momentum, $P$, is increased. The second derivative of the FMGS energy is discontinuous at a critical momentum, $P_{\rm crit}$, which is depicted by the solid black line in Fig.~\ref{fig:GS_PhaseDiagram}(a). We find that the FMGS energy has the functional form
\begin{equation}\label{eq:PolDispersion}
	E\left(P\right) = \begin{cases}
	f(P), & P < P_\textrm{crit}\\
	cP, &  P > P_\textrm{crit}
	\end{cases}
\end{equation}
and for small and intermediate interactions,
\begin{align}\label{eq:Pcrit}
f(P)= \frac{P^2}{2m^{*}}, \quad P_\textrm{crit}=m^{*} c.
\end{align}
where $m^{*}= \partial^2E/\partial P^2|_{P=0}$ is the effective mass of the polaron. The black dashed line in Fig.~\ref{fig:GS_PhaseDiagram}(a) depicts $P_\textrm{crit}$ from Eq.~\eqref{eq:Pcrit}, and we see that it matches the solid line for weak and intermediate interactions.

The polaron's group velocity, $v_\textrm{pol}=\partial E/\partial P$, is equal to the average velocity of the impurity in the FMGS, $v_\textrm{imp}^{\textrm{gs}}=\braket{\Psi_\textrm{gs}|\hat{\vec{P}}_\textrm{imp}|\Psi_\textrm{gs}}/m_{I}$, by the Hellmann-Feynman theorem~\cite{knap_flutter_signatures_2014}. Here, $\ket{\Psi_\textrm{gs}}$ is the FMGS and $\hat{\vec{P}}_\textrm{imp}$ is the impurity momentum operator. These velocities transition from an interaction-dependent subsonic value to the speed of sound when $P_\textrm{crit}$ is crossed, as illustrated in Fig.~\ref{fig:ObsComp}(a).


The impurity's momentum distribution function, $n_{\vec{P}_\mathrm{imp}}$, also exhibits a signature of the transition; its coherent, $\delta$-function part, with a weight given by the quasiparticle residue $Z = \abs{\braket{0|\Psi_\textrm{gs}}}^{2}$, vanishes when $P_\textrm{crit}$ is crossed, as shown in Fig.~\ref{fig:ObsComp}(b). Here, $\ket{0}$ is the plane-wave state corresponding to a non-interacting impurity immersed in a BEC. The shape of the residual, incoherent part of the momentum distribution function becomes sharply peaked at the critical momentum before broadening out again~\cite{KS_CherenkovLetter_SM}.

\begin{figure}[t!]
	\centering
	\includegraphics[clip, width=0.97\columnwidth]{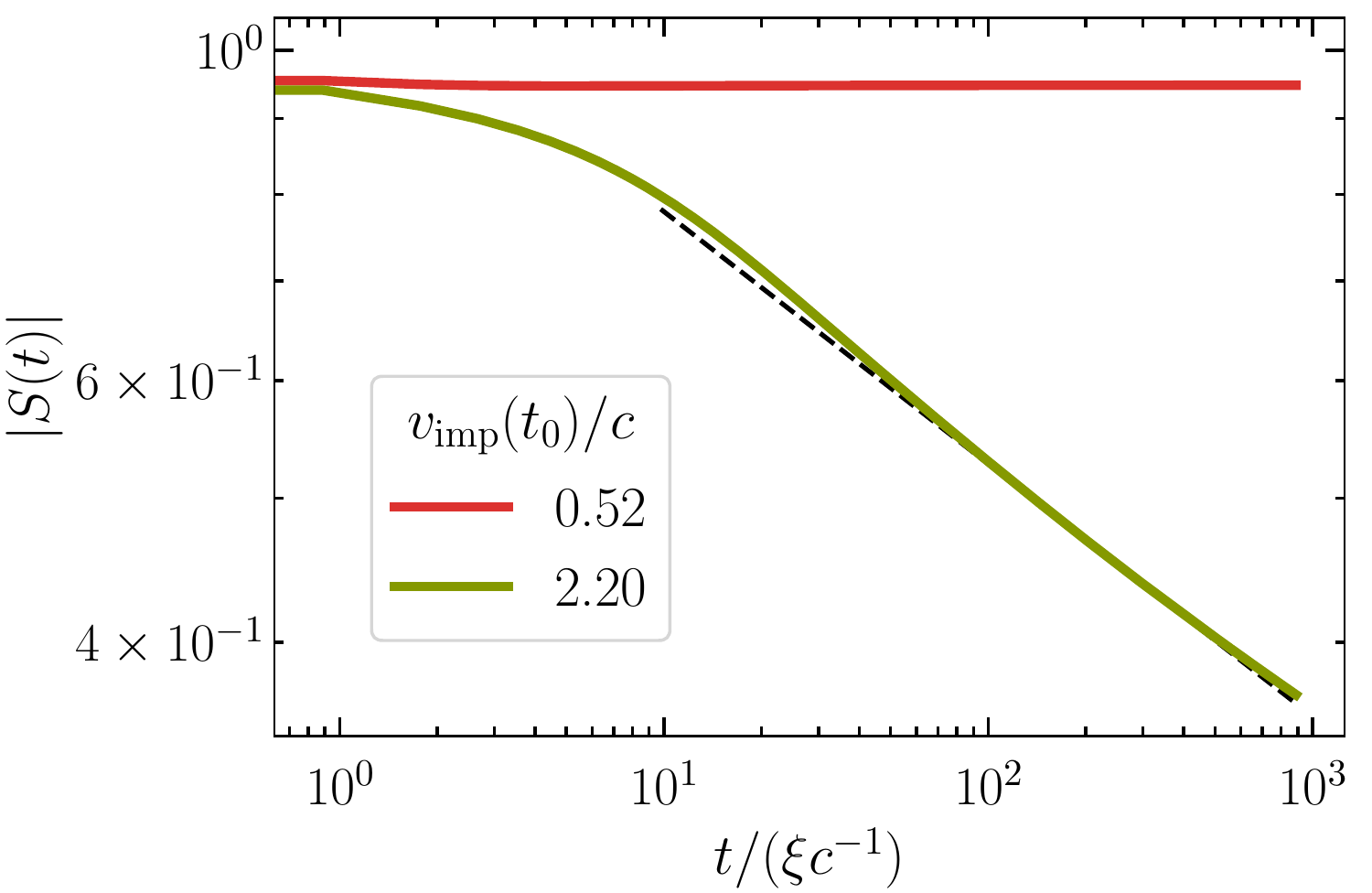}
	\caption{\textbf{Loschmidt echo dynamics.} The black dotted line is a visual guideline for the power-law decay of the Loschmidt echo in the Cherenkov regime. The impurity-boson scattering length is $a_{\rm IB}^{-1}=-8.92/\xi^{-1}$ and the impurity-boson mass ratio is $m_{I}/m_{B}=1$.}
	\label{fig:TimeDecay}
\end{figure}

The finite momentum transition also manifests in the dynamics of the system after the impurity is quenched from a non-interacting state to an interacting state. The orthogonality catastrophe defined by the vanishing $Z$ gives a natural motivation to examine the Loschmidt echo, $S(t)=e^{i\frac{P^2}{2m_{I}}t}\braket{0|\Psi(t)}$, which characterizes the dynamical transition and can be experimentally measured using Ramsey interferometry on the impurity atom~\cite{Knap2012,Cetina2016}. Here, $\ket{\Psi(t)}$ is the wavefunction of the interacting system at time $t$. Based on the equivalence of the long time limit of the Loschmidt echo, $|S(t_\infty)|$, and $Z$, shown analytically in Ref.~\cite{Shashi2014}, we examine the behavior of $S(t)$ for various initial momenta in the quench protocol and observe a dynamical transition; $S(t)$ remains finite for small initial system momenta, but has a power law decay $S\left(t\right) \sim t^{-\gamma}$ at long times for large initial momenta, as illustrated in Fig.~\ref{fig:TimeDecay}. The power-law behavior is reminiscent of the dynamical response in systems featuring orthogonality catastrophe, discussed for example in Ref.~\cite{Knap2012}. This form of the decay also confirms that the time evolution of the system is non-Markovian, as the use of the Markov approximation leads to an exponential decay of $S(t)$ and does not capture the dependence of $P_\textrm{crit}$ on the impurity-gas coupling strength~\cite{Lausch2018,Nielsen2019,Boyanovsky2019}.

We depict $|S(t_\infty)|$ in Fig.~\ref{fig:ObsComp}(d). Within our treatment of the system, its power law decay for large initial momenta corresponds to a logarithmic increase of total emitted phonon number at long times in the supersonic regime. The numerically extracted critical momenta for the long time limit of the dynamical protocol are depicted by the red diamonds in Fig.~\ref{fig:GS_PhaseDiagram}(a). These values differ from the extracted values of the FMGS transition at strong interactions; we believe these differences are due to the structure of our trial wavefunction, as discussed further in~\cite{KS_CherenkovLetter_SM}, and may be reconciled using a generalization of the wavefunction ansatz~\cite{Shi2017}.

Motivated by the fact that the impurity velocity acts as a signature of the FMGS transition, we also examine the dynamical impurity velocity $v_\textrm{imp}\left(t\right)=\braket{\Psi(t)|\hat{\vec{P}}_\textrm{imp}|\Psi(t)}/m_{I}$. We find that its long time limit, $v_\textrm{imp}\left(t_{\infty}\right)$, witnesses the same transition as $|S(t_\infty)|$. Specifically, we observe
\begin{equation}\label{eq:vimp_inf_asymp}
	v_\textrm{imp}\left(t_{\infty}\right) = \left\{ \begin{array}{ll}
	\mathrm{constant}<c, & \text{$P< P_{\rm crit}$}\\
	&\\
	c, & \text{$P>P_{\rm crit}$}
	\end{array} \right.
\end{equation}
as depicted in Fig.~\ref{fig:ObsComp}(c).


The density of atoms of the host liquid, $n_{a}\left(\vec{r},t\right)$, shows qualitative differences on the two sides of the dynamical transition. In the subsonic phase, the impurity generates a rounded wave in the host gas density which propagates outwards faster than the impurity's motion. In the Cherenkov phase, the impurity generates a shock wave and wake in the host liquid density, with this modulation traveling along with the impurity even at late times. We illustrate the typical Cherenkov and subsonic patterns in Fig.~\ref{fig:GS_PhaseDiagram}(b) and (c) respectively, where we plot the integrated density, $n_{a}\left(x,z,t\right)=\int dy\,n_{a}\left(\vec{r},t\right)$, at a specific time $t=40\,\xi/c$ in the frame of the impurity propagating in the $z$-direction. Both integrated densities, $n_{a}\left(x,z,t\right)$ and $n_{a}\left(z,t\right)=\int dx\,n_{a}\left(x,z,t\right)$, plotted in~\cite{KS_CherenkovLetter_SM}, are directly visible in experiments via absorption imaging.


In~\cite{KS_CherenkovLetter_SV}, we show the time-evolution of $n_{a}\left(x,z,t\right)$ for different interaction strengths and initial momenta. For interactions weak enough that an initially supersonic impurity decays towards  $c$ slowly, we observe the generation of a shock wave whose cone angle gradually becomes shallower as the impurity slows down. This dynamically changing cone angle, resulting from a changing Mach number $v_\textrm{imp}\left(t\right)/c$, occurs due to finite recoil in the system which is not present in the classical limit of an infinitely heavy impurity. For strong interactions, the average velocity of an impurity in the Cherenkov regime quickly decays to $c$ as it gets entangled with the host liquid. The impurity maintains a shallow shock wave and wake as it travels with constant average velocity $c$ through the host liquid. Initially supersonic impurities that are in the subsonic regime, however, stop generating a shock wave and wake as time progresses. The existence of a persistent shock wave at late times therefore depends on how much velocity an initially supersonic impurity loses as it gets entangled with the host liquid. This dependence of the observed Cherenkov effect on the velocity of the entangled impurity highlights the quantum nature of the transition arising from the impurity's finite mass and ensuing recoil.

If we make the boson-boson interaction in the host liquid weaker, the speed of sound decreases. In the limiting case of a non-interacting Bose gas, the system exhibits an orthogonality catastrophe where the FMGS quasiparticle residue will vanish for any choice of system parameters, a prediction that is consistent with Ref.~\cite{Guenther2020}.


\textit{Discussion}.---We find a quantum Cherenkov transition that exists for all impurity-boson mass ratios and interaction strengths, with the case of unequal mass ratios shown in~\cite{KS_CherenkovLetter_SM} . In the limit of infinite $m_{I}$, we recover the kinematic Landau criterion for an macroscopic classical obstacle propagating through a superfluid. Finite $m_{I}$ introduces the impurity's recoil energy as a relevant scale, thus making the quantum nature of the impurity important to the physics of the transition. For weak and intermediate interactions, the transition can be described in terms of mass renormalization by applying the Landau criterion to a polaron quasiparticle with effective mass $m^{*}$. For stronger interactions, however, we find that this description is no longer sufficient as $P_\textrm{crit}$ is larger than $m^{*}c$, and the momentum dependence of the FMGS energy, Eq.~\eqref{eq:PolDispersion}, can no longer be captured by just $m^{*}$. In contrast to the classical Cherenkov effect, a mobile quantum impurity injected into a quantum liquid expresses whether it is in the Cherenkov regime only at late times. Further details of the transition we find are discussed in Ref.~\cite{Seetharam2021_CherenkovLong}.

The existence of a finite momentum transition in other impurity systems, and whether the FMGS and dynamical manifestations of such a transition occur at the same point, can provide insight into how dynamical behavior can be classified. For example, the finite momentum behavior of an impurity interacting repulsively with a one-dimensional Fermi gas has been examined~\cite{Mathy2012, knap_flutter_signatures_2014,gamayun_protocol_impurity_17}. There, a FMGS transition exists as momentum is increased, while a dynamical quench protocol similar to the one in our work exhibits a crossover. Specifically, in the one-dimensional gas studied in Ref.~\cite{gamayun_protocol_impurity_17}, the behavior of the post-quench long time limit of the average impurity velocity, $v_\textrm{imp}(t_\infty)$, is a smooth and non-monotonous function of the initial velocity $v_\textrm{imp}(t_{0})$; it never reaches the speed of sound in the gas. Furthermore, $v_\textrm{imp}(t_\infty)\to 0$ as $v_\textrm{imp}(t_{0})\to\infty$. This behavior may be associated with the prevailing role of a few initial central collisions between the impurity and host particles in the one-dimensional gas; such collisions take away virtually all momentum from the impurity. In contrast, for the three-dimensional gas studied in our work, small-angle collisions play an important role, thus ensuring that the impurity may have $ v_\textrm{imp}(t_\infty)$ remain far from zero for any arbitrarily high initial impurity velocity as shown in Fig.~\ref{fig:ObsComp}(c). Two-dimensional gases may have collisions in-between these two cases, and therefore we have no \textit{a priori} conclusions about the behaviour of $v_\textrm{imp}(t_\infty)$ for large initial velocities in these systems.

Experimentally, the transition we observe can be detected through time-of-flight imaging of a dilute impurity gas immersed in a BEC. We have established that the width of the impurity's momentum distribution acts a signature of the transition in the FMGS. Alternatively, Ramsey interferometry or RF spectroscopy, as detailed in Ref.~\cite{Cetina2016}, can be used to detect the dynamical transition as the onset of a power-law decay in the Loschmidt echo signifies the critical point. Absorption imaging of the density distribution of host atoms would directly reveal the polaron shock wave and wake of sufficiently fast impurities. We assume a homogeneous BEC in the vicinity of the impurity, which can be realized using a sufficiently large box trap. The dynamical quantum Cherenkov transition we uncover provides a route to experimentally pinpoint the characteristic dynamical properties of impurities immersed in quantum liquids, which would grant insight into the far-from-equilibrium behavior of quantum many-body systems.


\textit{Acknowledgements}.---We thank Zoe Yan, Carsten Robens, Yiqi Ni, Richard Schmidt, Tao Shi, Achim Rosch, Martin Zwierlein, and Immanuel Bloch for helpful discussions. This research was conducted with Government support under and awarded by DoD, Air Force Office of Scientific Research, National Defense Science and Engineering Graduate (NDSEG) Fellowship, 32 CFR 168a. F. G. acknowledges support by the Deutsche Forschungsgemeinschaft (DFG, German Research Foundation) under Germany's Excellence Strategy -- EXC-2111 — 390814868. The work of M. B. Z. is supported by Grant No. ANR-
16-CE91-0009-01 and CNRS grant PICS06738. E. D. acknowledges support by the Harvard-MIT Center of Ultracold Atoms, ARO grant number W911NF-20-1-0163, and NSF EAGER-QAC-QSA.



%
\clearpage
\pagebreak
\onecolumngrid

\renewcommand{\theequation}{S.\arabic{equation}}
\renewcommand{\thesection}{S\arabic{section}}
\renewcommand{\thefigure}{S\arabic{figure}}

\setcounter{equation}{0}
\setcounter{figure}{0}

\section{\large Supplementary Material}


\section{Impurity momentum magnitude distribution (FMGS)}\label{sec:impDist}

We examine the distribution of the impurity's momentum magnitude
\begin{equation}\label{eq:impMomMagDist}
n_{\rm imp}(p)=\sum_{\vec{p'}}\delta\left(\abs{\vec{p'}}-p\right)n_{\rm imp}(\vec{p'})
\end{equation}
in the finite momentum ground state (FMGS), where
\begin{align}
n_{\rm imp}(\vec{p}) &= \left\langle \delta\left(\hat{\vec{P}}_{\rm imp}-\vec{p}\right)  \right\rangle\label{eq:ImpMomDist}\\
& = Z\delta\left(\vec{P} - \vec{p}\right) + \tilde{n}_{\rm imp}(\vec{p})\nonumber.
\end{align}
The three-dimensional momentum distribution, Eq.~\eqref{eq:ImpMomDist}, is comprised of a $\delta$-peak with magnitude given by the quasiparticle residue, $Z$, and an incoherent part, $\tilde{n}_{\textrm{imp}}(\vec{p})$. The momentum magnitude distribution, Eq.~\eqref{eq:impMomMagDist}, is therefore also comprised of a $\delta$-peak with magnitude $Z$ and an incoherent part.

\begin{figure*}[h!]
	\centering
	\includegraphics[width=0.99\textwidth]{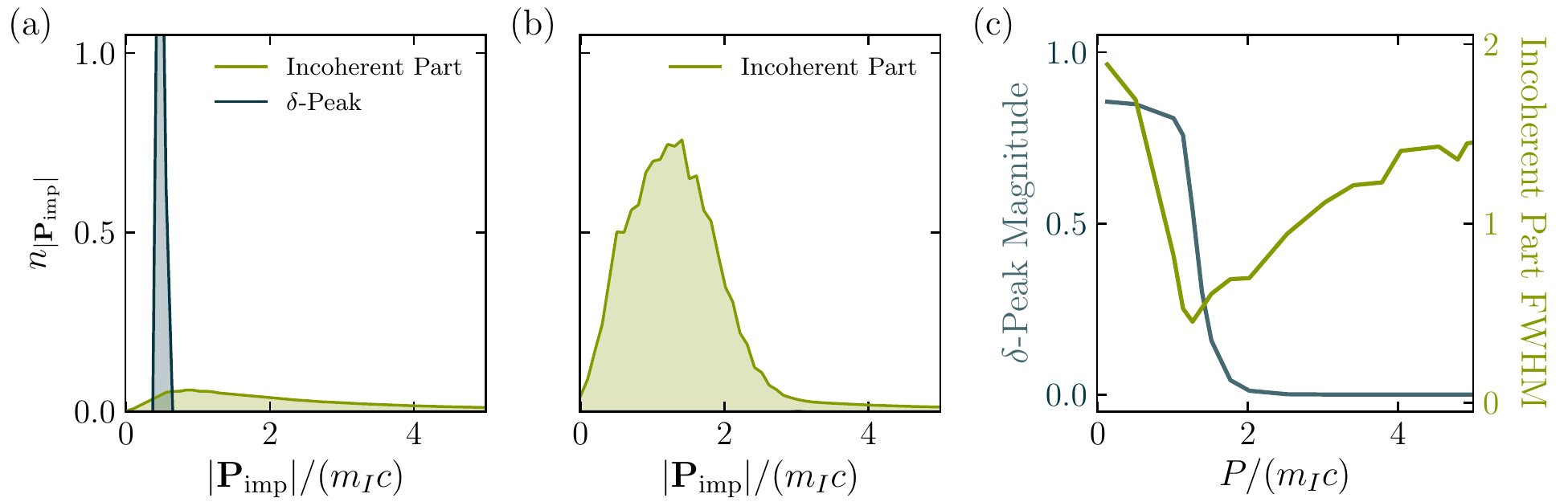}
	\caption{\textbf{Distribution of impurity's momentum magnitude in FMGS}. \textbf{(a)} Subsonic regime ($P=0.5m_{I}c$). \textbf{(b)} Cherenkov regime ($P=3.0m_{I}c$). Note that we have introduced Gaussian broadening of the $\delta$-peak in the distributions by hand; in reality we get a $\delta$-peak at $|\vec{P}_\textrm{imp}|=P$ with a magnitude corresponding to the quasiparticle residue $Z$. \textbf{(c)} Magnitude of $\delta$-peak, $Z$, and full width at half maximum (FWHM) of incoherent part of the distribution for different values of total system momentum $P$. The impurity-boson scattering length is $a_{\rm IB}^{-1}=-4.46/\xi^{-1}$ and the impurity-boson mass ratio is $m_{I}/m_{B}=1$.}
	\label{fig:FMGS_ImpDist}
\end{figure*}

In Fig.~\ref{fig:FMGS_ImpDist}(a) and (b), we illustrate examples of $n_{\rm imp}(p)$ in the subsonic and Cherenkov regimes respectively. In Fig.~\ref{fig:FMGS_ImpDist}(c), we plot the magnitude of the $\delta$-peak and full width at half maximum of the incoherent part of the distribution as a function of total system momentum. We see that the magnitude of the $\delta$-peak, given by the quasiparticle residue, is finite at low momenta corresponding to the subsonic regime, and then sharply decays to zero upon crossing the Cherenkov transition. At the transition point, the width of the incoherent part of the distribution becomes much narrower. This feature can be used to diagnose the FMGS transition in experiments via time of flight imaging.

\pagebreak

\section{Host liquid density distribution (dynamics)}\label{sec:intDensity}

\begin{figure*}[h!]
	\centering
	\includegraphics[width=0.99\textwidth]{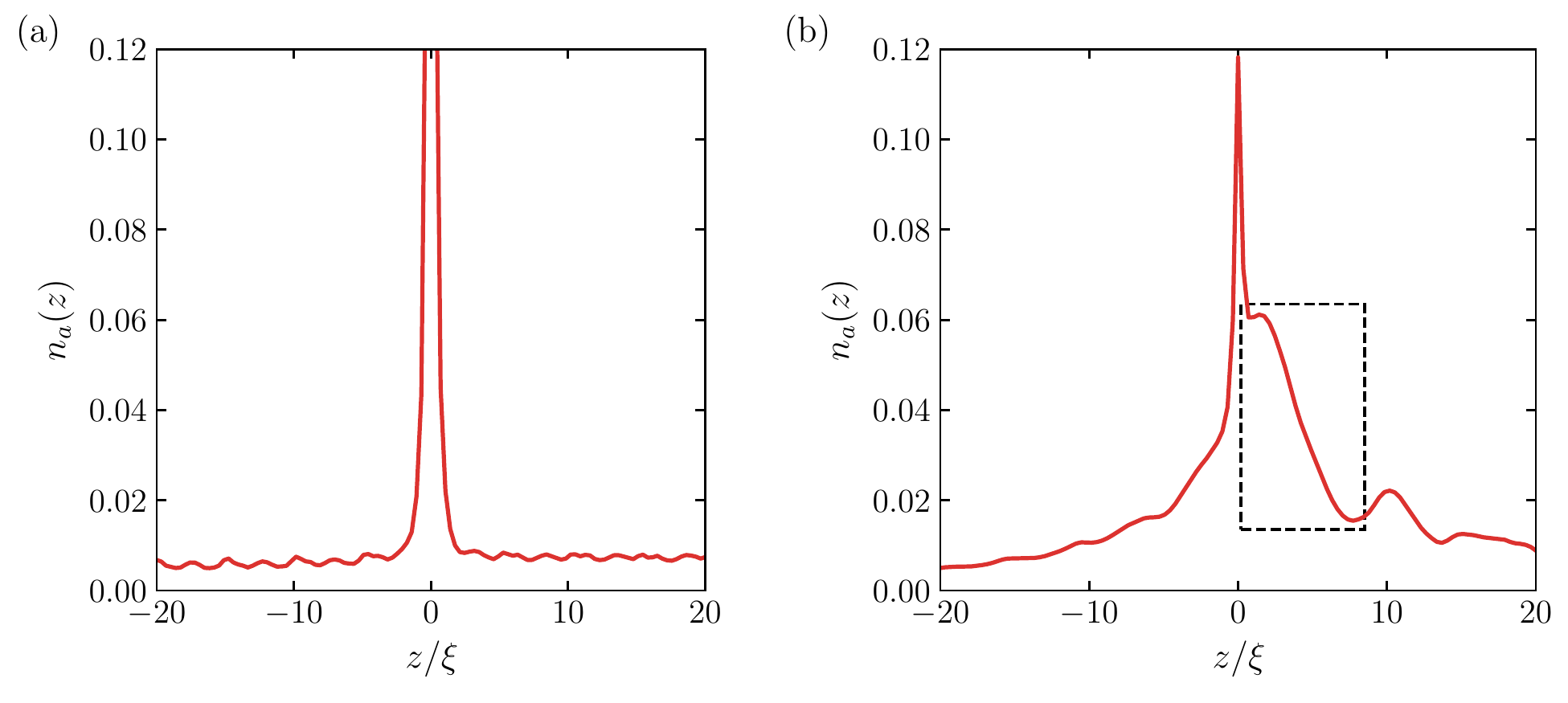}
	\caption{\textbf{Integrated host liquid density distribution in direction of impurity propagation}. \textbf{(a)} Subsonic regime ($P=0.5m_{I}c$). \textbf{(b)} Cherenkov regime ($P=3.0m_{I}c$).  The dashed black rectangle highlights the shock wave in front of the impurity. The distributions are plotted at time $t=40\,\xi/c$ in the frame of the impurity. The impurity-boson scattering length is $a_{\rm IB}^{-1}=-4.46/\xi^{-1}$ and the impurity-boson mass ratio is $m_{I}/m_{B}=1$.}
	\label{fig:Dyn_densityDist}
\end{figure*}

Here, we integrate the real space density distribution of the host liquid plotted in Fig.1(b) and (c) of the main text across the direction perpendicular to the impurity's motion. Specifically, in Fig.~\ref{fig:Dyn_densityDist}, we plot $n_{a}\left(z,t\right)=\int dx\,dy\,n_{a}\left(\vec{r},t\right)$ at a specific time $t=40\,\xi/c$ in the frame of the impurity propagating in the $z$-direction. Panels (a) and (b) in Fig.~\ref{fig:Dyn_densityDist} are typical patterns of the density in the subsonic and Cherenkov regimes respectively. We see a peak at the origin in both panels corresponding to the position of the impurity. In Fig.~\ref{fig:Dyn_densityDist}(b), representing the density in the Cherenkov regime, we see a shock wave followed by the first oscillation of the wake. The shock wave appears at approximately one healing length in front of the impurity ($z=\xi$).

\section{Phase transition at different mass ratios (dynamics)}\label{sec:massRat}

\begin{figure*}[h!]
	\centering
	\includegraphics[width=0.8\textwidth]{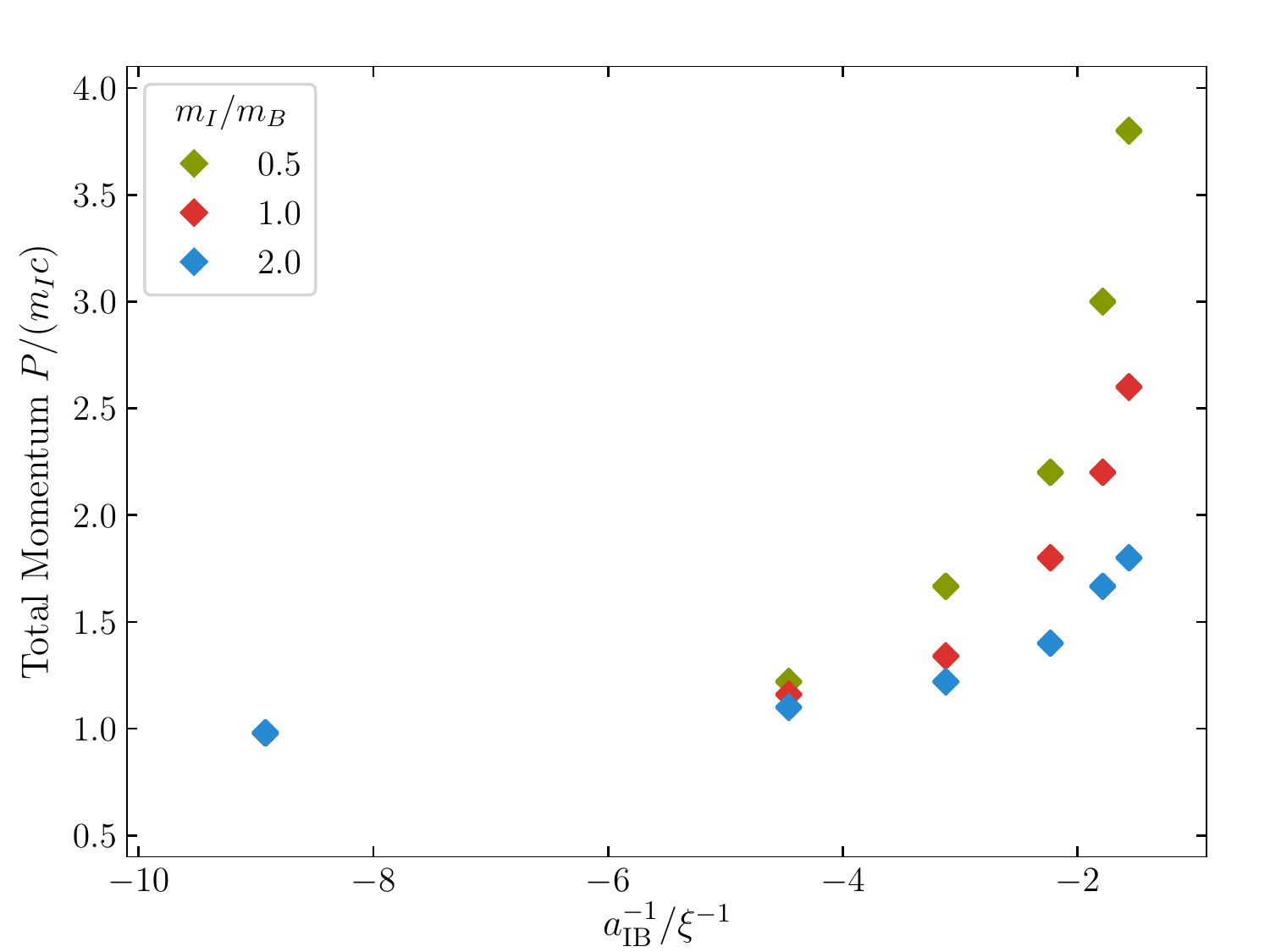}
	\caption{\textbf{Dynamical quantum Cherenkov transition for different impurity-boson mass ratios}.}
	\label{fig:Dyn_massRat}
\end{figure*}

In Fig.~\ref{fig:Dyn_massRat}, we show the dynamical transition points $P_\textrm{crit}$, depicted by the red diamonds in Fig.~1(a) of the main text, for various impurity-boson mass ratios. We see that for a fixed interaction strength, heavier impurities have a smaller critical momentum closer to $m_{I}c$; interactions must be stronger to appreciably renormalize the mass of heavy impurities. In the limit of an infinitely heavy impurity and finite interaction strength, the interactions have a negligible effect and the transition occurs at the point of the classical Cherenkov transition, $P_\textrm{crit}\rightarrow m_{I}c$.

\section{Methods}\label{sec:Model}

The theoretical treatment follows the approach in Ref.~\cite{Shchadilova2016}. We derive equations of motion for the system using a non-Gaussian variational approach, where we combine a Lee-Low-Pines rotation of the system along with a variational coherent state ansatz in the rotated frame~\cite{Shchadilova2016,Drescher2019}. The Lee-Low-Pines transformation allows the variational coherent state to capture most relevant correlations of all orders between the impurity and boson degrees of freedom~\cite{Lee1953}. We compute the FMGS and quench dynamics properties of the system as a function of $P$ and $a_\textrm{IB}$ by running the equations of motion in imaginary time and real time respectively. The Hamiltonian used to model an impurity immersed in a weakly interacting BEC of ultracold atoms is described below. 

We start with a microscopic Hamiltonian consisting of the kinetic energy of the impurity and host gas, contact interactions between atoms of the host gas, and contact interactions between the impurity and host gas atoms. The system is in a three-dimensional box, $d=3$, of length $L$ in each dimension, and periodic boundary conditions are imposed. Standard Bogoliubov theory is then applied to expand the host Bose gas around its condensate mode, with the impurity coupling to the condensate and Bogoliubov excitations on top of the condensate~\cite{pitaevskii_book_BEC}. Now we take our model, $\hat{H}$, defined in the laboratory reference frame and write it in the mobile impurity reference frame. These reference frames are connected with the Lee-Low-Pines transformation,
\begin{equation}
\hat{\mathcal{S}}= e^{i\hat{\vec{R}}_\textrm{imp}\hat{\vec{P}}_\textrm{ph}},
\end{equation}
named after the original work~\cite{Lee1953}. Here, $\hat{\vec{R}}_\textrm{imp}$ is the position operator of the impurity and $\hat{\vec{P}}_\textrm{ph}$ is the total phonon momentum, where we refer to Bogoliubov excitations of all wavelengths as phonons. Any operator $\hat{\mathcal{O}}$ defined in the laboratory frame takes the form
\begin{equation}
\hat{\mathcal{O}}_\textrm{LLP} = \hat{\mathcal{S}} \hat{\mathcal{O}} \hat{\mathcal{S}}^\dagger
\end{equation}
in the mobile impurity reference frame.

The total momentum of the system
\begin{equation}\label{eq:totMom}
\hat{\vec{P}} = \hat{\vec{P}}_{\rm imp} + \hat{\vec{P}}_{\rm ph}
\end{equation}
is conserved.  In the above expression, the impurity momentum $\hat{\vec{P}}_{\rm imp}$ and total phonon momentum $\hat{\vec{P}}_{\rm ph}$ are given as
\begin{align}\label{eq:impPhonMom}
\hat{\vec{P}}_{\rm imp} & = \sum_{\vec{k}} \vec{k} \hat d^\dagger_{\vec{k}}\hat d_{\vec{k}}\\
\hat{\vec{P}}_{\rm ph} & = \sum_{\vec{k}} \vec{k} \hat b^\dagger_{\vec{k}}\hat b_{\vec{k}}.\label{eq:avgPhononMom}
\end{align}
where $\hat{d}^\dagger_{\vec{k}}$ ($\hat{d}_{\vec{k}}$) is the impurity creation (annihilation) operator and $\hat{b}^\dagger_{\vec{k}}$ ($\hat{b}_{\vec{k}}$) is the same for a Bogoliubov excitation.

The final Hamiltonian $\hat{H}_{\rm LLP}$ after the transformation reads:
\begin{equation}\label{eq:HLLP}
\hat{H}_{\rm LLP} \equiv \hat{\mathcal{S}} \hat{H}\hat{\mathcal{S}}^{\dagger}
=\hat{H}_{0, \rm LLP}+\hat{H}_{\rm IB, \rm LLP}
\end{equation}
where
\begin{equation}\label{eq:H_0LLP}
\hat{H}_{0, \rm LLP} =\sum_{\vec{k}}\omega_{\vec{k}}\hat{b}_{\vec{k}}^{\dagger}\hat{b}_{\vec{k}} + \frac{1}{2m_{I}}\left(\hat{\vec{P}}_{\rm imp}-\sum_{\vec{k}}\vec{k}\hat{b}_{\vec{k}}^{\dagger}\hat{b}_{\vec{k}}\right)^{2}
\end{equation}
and
\begin{align} \label{eq:H_IBLLP}
\hat{H}_{\rm IB, \rm LLP}  &=g_{\rm IB}n_{0} + g_{\rm IB}\sqrt{n_{0}}\frac{1}{\sqrt{L^{d}}}\sum_{\vec{k}\neq0}W_{\vec{k}}\left( \hat{b}_{\vec{k}}+\hat{b}_{-\vec{k}}^{\dagger}\right) \\
&+g_{\rm IB}\frac{1}{L^{d}}\sum_{\vec{k}\neq0,\vec{k'}\neq0}V_{\vec{k},\vec{k'}}^{\left(1\right)}\hat{b}_{\vec{k}}^{\dagger}\hat{b}_{\vec{k}'}
+\frac{1}{2}g_{\rm IB}\frac{1}{L^{d}}\sum_{\vec{k}\neq0,\vec{k'}\neq0}V_{\vec{k},\vec{k}'}^{\left(2\right)}\left(\hat{b}_{-\vec{k}}\hat{b}_{\vec{k'}}+\hat{b}_{\vec{k}}^{\dagger}\hat{b}_{-\vec{k}'}^{\dagger}\right).
\end{align}
Here,
\begin{equation} \label{eq:omegaBEC}
\omega_{\vec{k}}=\sqrt{\frac{\vec{k}^2}{2m_{B}}(\frac{\vec{k}^2}{2m_{B}}+2g_\mathrm{BB}n_{0})}.
\end{equation}
is the dispersion of Bogoliubov excitations, $m_{B}$ is the mass of bosonic atoms comprising the host gas, $g_\textrm{IB}$ and $g_\textrm{BB}$ are the impurity-boson and boson-boson contact interaction strengths, $n_{0}$ is the condensate density, and the interaction vertices satisfy the relations 
\begin{align}
W_{\vec{k}}&= \sqrt{\frac{\vec{k}^2}{2m_{I}\omega_{\vec{k}}}}\\
V_{\vec{k}\vec{k'}}^{(1)} \pm V^{(2)}_{\vec{k} \vec{k'}} &= \left( W_{\vec{k}} W_{\vec{k'}}\right)^{\pm 1}
\end{align}
where $m_{I}$ is the impurity mass. Note that the lab frame impurity momentum operator $\hat{\vec{P}}_{\rm imp}$ that appears in Eq.~\eqref{eq:H_0LLP} commutes with the Hamiltonian $\hat{H}_{\rm LLP}$ in the Lee-Low-Pines frame. We recognize it as the total momentum $\hat{\vec{P}}$ that commutes with the lab frame Hamiltonian $\hat{H}$. Therefore, we can replace $\hat{\vec{P}}_{\rm imp}$ with the quantum number $\vec{P}$; dynamics under this model will decouple into sectors indexed by $\vec{P}$. As the system is spherically symmetric, there is no preferred orientation of $\vec{P}$ and hence physical quantities will only depend on the magnitude $P=\abs{\vec{P}}$. We derive equations of motion for the system using the time-dependent variational principle and a multimode coherent state ansatz, which amounts to replacing the operators $\hat{b}^\dagger_{\vec{k}}$ ($\hat{b}_{\vec{k}}$) in Eq.~\eqref{eq:HLLP} with their expectation values. Specfically, we use a trial wavefunction of the form
\begin{equation} \label{eq:WF}
\ket{\Psi_\textrm{coh}(t)} = e^{\sum_{\vec k} \beta_{\vec k}(t) \hat{b}_{\vec k}^\dag - \rm H.c. } \ket{0}
\end{equation}
which is parameterized by the complex numbers $\{\beta_\vec{k}\}$. This choice allows for straightforward calculation of the equations of motion, which in turn are simple enough to numerically compute on large grids which approximate the continuum $\vec k$ limit well. The wavefunction structure capture the essential physics of the polaron system over a range of interaction strengths as it allows a large occupation of phonon modes, $\hat{b}_{\vec k}$, with the number of these excitations increasing for strong interactions~\cite{Shashi2014, Shchadilova2016}. Entanglement between the impurity and host atoms is captured as these phonon excitations are defined in the Lee-Low-Pines frame. However, the wavefunction Eq.~\eqref{eq:WF} does not explicitly account for correlations between different excitations, which are only implicitly accounted for when the dynamics of the true many-body wavefunction is projected onto the variational manifold.  At strong interactions, these correlations become relevant as the number of excitations becomes sizable; this fact may account for the numerical discrepency in the critical momenta between the FMGS and dynamical protocols as see in Fig.~1(a) of the main text. Promoting the trial wavefunction Eq.~\eqref{eq:WF} to a full Gaussian state~\cite{Shi2017}, amounting to including terms quadratic in $\hat{b}_{\vec k}$ in the exponential, may resolve this quantitative discrepency.

\end{document}